\documentclass[]{article}

\usepackage{graphicx}

\begin{document}

\begin{center}

{\bf Prospects for a Nuclear Optical Frequency Standard based on Thorium-229}\\

\bigskip

{\sc E. Peik, K. Zimmermann, M. Okhapkin, Chr. Tamm}

\bigskip

{\it Physikalisch-Technische Bundesanstalt,\\
Bundesallee 100, 38116 Braunschweig, Germany}

\end{center}

\vspace{1cm}

\begin{abstract}
The 7.6-eV-isomer of Thorium-229 offers the opportunity to perform high resolution laser spectroscopy of a nuclear transition. We give a brief review of the investigations of this isomer. The nuclear resonance connecting ground state and isomer may be used as the reference of an optical clock of very high accuracy using trapped and laser-cooled thorium  ions, or in a compact solid-state optical frequency standard of high stability.
\end{abstract}

\vspace{1cm}

\bigskip
\noindent
{\bf 1. The low-lying isomer of $^{229}$Th}
\medskip

Thorium-229 seems to be a unique system in nuclear physics in that it possesses the only known isomer with an excitation energy in the range of optical photon energies and in the range of outer-shell electronic transitions. $^{229}$Th is part of the decay chain of $^{233}$U and undergoes $\alpha$-decay with a halflife of 7880 years. Its energy level structure was studied by the group of C. W. Reich at the Idaho National Engineering Laboratory since the 1970s, mainly relying on spectroscopy of the $\gamma$-radiation emitted after the $\alpha$-decay of $^{233}$U \cite{kroger,helmer}. It was noted that the lower energy levels belong to two rotational bands whose band heads must be very close, one being the ground state, the other the isomer. Evaluating several $\gamma$-decay cascades, the value $(3.5\pm 1.0)$~eV was obtained for the isomer energy \cite{helmer}. Further studies confirmed and extended the knowledge on the overall nuclear level scheme of $^{229}$Th \cite{gulda,barci,ruchowska} and also supported the value for the isomer excitation energy \cite{barci}. The isomer may decay to the ground state under the emission of magnetic dipole radiation with an estimated lifetime of a few 1000 s \cite{dykhne,ruchowska} in an isolated nucleus.     

These results on the extremely low excitation energy of the isomer inspired a number of theoretical studies, investigating the decay modes of the isomer in different chemical surroundings and possible ways of exciting it with radiation (see Refs. \cite{matinyan,tkalya} for reviews). On the experimental side, two false optical detections of the decay of the isomer were reported \cite{irwin,richardson}, but it was quickly clarified that the observed light was luminescence induced by the background of $\alpha$-radiation \cite{utter,shaw}. Fluorescence experiments with radioactive samples and ultraviolet light sources may be severely affected by thermoluminescence, Cherenkov radiation etc. \cite{young,peik1}. 
All further attempts at a direct observation of the optical transition connecting ground state and isomer failed. 

A reanalysis of the data presented in Ref. \cite{helmer} indicated a possible shift to higher energies by about +2 eV \cite{helene}. New experimental data on the transition energy became available only after a group at LLNL used a high resolution $\gamma$-spectrometer and measured two decay cascades very precisely. The result on the transition energy is $(7.6\pm0.5)$~eV \cite{beck}, placing the transition in the vacuum ultraviolet at about 160 nm. Radiation at this wavelength is not transmitted through air, water or optical glasses, which explained the failure of  most of the previous attempts to detect the radiation emitted in the decay of the isomer. The energy of 7.6~eV is higher than the ionization potential of the Th atom but lower than that of Th$^+$. This opens internal conversion as an alternative decay channel in neutral thorium. 

\begin{figure}
\begin{center}
\includegraphics[width=5cm]{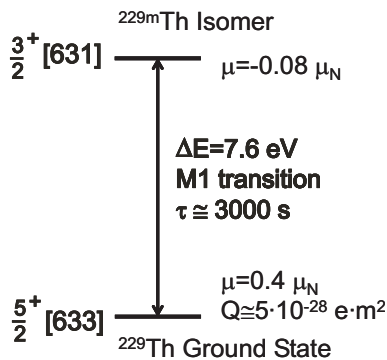}
\end{center}
\caption{The ground state and lowest excited state of the $^{229}$Th nucleus with the level classification in the Nilsson model, radiative lifetime \cite{ruchowska,dykhne} for the magnetic dipole transition, magnetic moments in nuclear magnetons and quadrupole moment of the ground state \cite{gerstenkorn,dykhne}.}
\end{figure}

We have extended the search for light emission from $^{229}$Th isomers produced in the $\alpha$-decay of U-233 using photodetectors with  sensitivity in the VUV range. Light emission from a solid $^{233}$U-nitrate sample as well as from freshly produced $^{229}$Th recoil nuclei in a catcher foil was investigated -- without seeing the expected signals \cite{zimmermann}. In the experiment with $^{233}$U a spectrometer was used but the noise level on the detector was too high to reach the required sensitivity. The experiment with the recoil nuclei showed negligible background and the negative outcome may be interpreted as a sign for a rapid nonradiative decay of the isomer in the solid.

The expected transition wavelength is in a part of the VUV range that is accessible by frequency upconversion of narrow-bandwidth continuous and phase-coherent femtosecond laser sources. Thus it appears possible to measure the nuclear energy level splitting of $^{229}$Th with the precision afforded by high-resolution laser spectroscopy and optical frequency metrology. 
Figure 1 summarizes the relevant spectroscopic information that is available today.

\bigskip
\noindent
{\bf 2. Nuclear optical clock with trapped ions}
\medskip

Nuclear transition frequencies are much more stable against external perturbations than transition frequencies of the electron shell because the characteristic nuclear dimensions are small compared to the atomic dimension. Therefore nuclear transitions are attractive as highly accurate frequency references with small field-induced shifts \cite{peik2}. 
Apart from motional frequency shifts that can be well controlled e.g. in laser-cooled trapped ions, the interaction with ambient electric or magnetic fields usually is the dominant source of systematic uncertainty in optical frequency standards.
Estimates on the magnitude of systematic frequency shifts must, however, also consider the coupling of the nuclear and electronic energy level systems through the Coulomb and hyperfine interactions. For external electric field gradients the electron shell may actually lead to an enhancement at the nucleus (Sternheimer anti-shielding). It will therefore be important to select a suitable electronic state for the nuclear excitation.   

In order to illustrate the role of hyperfine interactions in nuclear spectroscopy of an isolated atom or ion, 
let us consider the Zeeman and Stark shifts of the nuclear transition frequency.
In an $LS$ coupling scheme the eigenstates of the coupled electronic and nuclear system are characterised by sets of quantum numbers $|\alpha,I;\beta,L,S,J;F,m_F\rangle$, where $I$ denotes the nuclear spin, $L,S,J$ the orbital, spin and total electronic angular momenta, and $F$ and $m_F$ the total atomic angular momentum and its orientation.
$\alpha$ and $\beta$ label the nuclear and electronic configurations. 
In the nuclear transition, the nuclear and total angular momentum quantum numbers ($\alpha, I, F, m_F$) can change, while the purely electronic quantum numbers ($\beta,L,S,J$) remain constant.
The nuclear transition frequency is independent of all mechanisms that produce level
shifts depending only on the electronic quantum numbers ($\beta,L,S,J$), because these do not change and consequently the upper and the lower state of the transition are affected in the same way. This applies to the
scalar part of the quadratic Stark effect, which typically is the dominant mechanism for the shift of electronic transition frequencies due to static electric fields, electromagnetic radiation, and collisions.
The observed nuclear transition frequency is however shifted by the hyperfine Stark shift, which depends on $F$ and $m_F$, and has been studied in microwave atomic clocks. In the optical frequency range, a relative magnitude of typically $10^{-19}$ may be expected for the hyperfine Stark shift caused by the $\approx 10$~V/cm room temperature blackbody radiation field. 
In order to avoid the influence of the linear Zeeman effect, an electronic state can be chosen such that $F$ is an integer. In this case a Zeeman component $m_F=0 \rightarrow 0$ is available, that shows only a small quadratic Zeeman effect around zero magnetic field. Since this shift depends   similarly on the electronic and the nuclear g-factor its magnitude will be comparable to those in other atomic clocks. 
Further field dependent shifts may arise from the tensor part of the quadratic Stark effect and from the quadrupole interaction between the atomic quadrupole moment and electric field gradients. Both these shifts can be expressed as a product of $J$-dependent and $F$-dependent terms  and vanish if either $J<1$ or $F<1$.

From these general considerations it can be seen that for every radiative nuclear transition,
an electronic state can be selected which makes the hyperfine coupled nuclear transition frequency immune against the linear Zeeman effect and the quadratic Stark effect as well as the quadrupole shift. 
For electronic transitions, this combination of advantageous features can not be obtained.
Since the selected electronic state has to be stable or at least long-lived, the choice could be made among the ground states of the differently charged positive ions of the element in question. 
In the case of a half integer nuclear spin (like in $^{229}$Th),
the optimal electronic states are $^2S_{1/2}$ or $^2P_{1/2}$, and for an integer nuclear spin the states 
$^1S_0$ or $^3P_0$ fulfill all criteria.

For a high precision nuclear clock, the case of trapped $^{229}$Th$^{3+}$-ions seems to be especially promising  \cite{peik2} because its electronic level structure is suitable for laser cooling. The sensitive detection of excitation to the isomeric state will be possible using a double resonance scheme that probes the hyperfine structure of a transition in the electron shell. No electric dipole transitions originate from the electronic ground state of Th$^{3+}$ in the range of 1.8 -- 15 eV so that resonant coupling between electronic and nuclear excitations is not expected to play an important role for the decay of the isomeric state.
The $5f~^2F_{5/2}$ ground state of Th$^{3+}$ does not fulfill the condition $J<1$ for elimination of the tensor Stark effect and the quadrupole shift, but a metastable $7s~^2S_{1/2}$ state of lifetime $\approx 1$~s is also available. Alternatively, the method of quantum logic spectroscopy \cite{schmidt} with an auxiliary ion may be applied to other charge states of $^{229}$Th that can not be laser-cooled directly. 

\bigskip
\noindent
{\bf 3. A solid-state nuclear frequency standard}
\medskip

A nuclear transition may also provide a resonance with very high resolution if the nuclei are embedded in a solid, as it is observed in M\"o{\ss}bauer spectroscopy. Thorium-229 opens the possibility to perform optical  M\"o{\ss}bauer spectroscopy using a laser as a tuneable, coherent source of radiation. This may provide a compact and simple reference for an optical frequency standard with performance much superior to what is available in simple atomic systems like vapor cells \cite{peik2,hudson}.  

The host crystal should be transparent at the nuclear resonance wavelength $\lambda_0$, a criterion that is fulfilled by a number of candidates like the fluorides of the alkaline earths. It would be only lightly doped with $^{229}$Th. If the broadening is dominantly homogeneous one nucleus per $\lambda_0^3$ may be used in order to avoid strong radiation trapping. Still, this would allow to handle $10^{11}$ nuclei in a cube of 1~mm dimension. With this number of nuclei direct fluorescence detection of the resonance radiation would be possible even if the resonant scattering rate is only of the order $10^{-4}$/s per nucleus.

The uncertainty budget of such a solid-state nuclear clock will be quite different from that of a realization with trapped ions considered above.
The crystal field shifts of the nuclear resonance frequency will be dominantly due to electric fields if a diamagnetic host is used. In insulators with high bandgap like fluorides rather high internal electric fields and field gradients will be found. The electron charge density at the position of the nucleus will lead to the isomer shift $\Delta f_{iso}=Ze^2\rho_0\langle r^2\rangle/(h\epsilon_0)$, where $Ze$ is the nuclear charge, $\rho_0$ the electron density at the nucleus and $\langle r^2\rangle$ the mean squared nuclear charge radius. The contribution of a 7s electron in thorium would shift the nuclear ground state by $\Delta f_{iso}\approx1$~GHz with respect to its energy in a bare nucleus. An electric field gradient will produce a quadrupole shift that may be of comparable magnitude: In the tetragonal crystal ThB$_4$, for example, the field gradient along the principal axis is about $5\times10^{21}$~V/m$^2$. Coupling to the ground state quadrupole moment of $^{229}$Th of $5\times 10^{-28}$~m$^2$ would produce a quadrupole shift of 0.6 GHz. The field gradient can be avoided in a crystal lattice of higher symmetry, like a cubic one.   

Both these shifts would be of less concern if they would be constant, which would only be the case if the positions of all charges in the lattice would be rigorously fixed. Thermal motion, however, will lead to a temperature-dependent broadening and shift of the line, where the line shape will depend on phonon frequencies and correlation times. Much information on these effects has been obtained in conventional ($\gamma$-ray) M\"o{\ss}bauer spectroscopy already. While the relativistic Doppler shift will lead to a temperature dependent relative frequency shift of about  $10^{-15}/$K, the temperature dependence of the crystal field will critically depend on the choice of crystal host and may be significantly bigger. For a solid state nuclear clock of high accuracy (beyond $10^{-15}$) the temperature dependence may be eliminated if the crystal is cryogenically cooled to well below the Debye temperature, so that the influence of phonons is effectively frozen out.

\bigskip
\noindent
{\bf 4. Conclusion}
\medskip

Nuclear laser spectroscopy of $^{229}$Th seems to offer great potential for frequency metrology and promises to open a new field of research at the borderline between nuclear and atomic physics, shedding new light on familiar phenomena like nuclear radiative decay or hyperfine interactions. It may allow improved tests of fundamental physics, as it was recently shown that the resonance frequency would be the most sensitive probe in the search for temporal variations of the fundamental coupling constants \cite{flambaum1, flambaum2}.       

\newpage
        
\noindent
{\bf Acknowledgments}
\medskip

This work is supported by DFG through SFB 407 and the cluster of excellence QUEST.

\end{document}